# WHAT'S WRONG WITH EINSTEIN'S 1927 HIDDEN-VARIABLE INTERPRETATION OF QUANTUM MECHANICS?*

Peter Holland

Green College
University of Oxford
Woodstock Road
Oxford OX2 6HG
England

peter.holland@green.ox.ac.uk

Einstein's unpublished 1927 deterministic trajectory interpretation of quantum mechanics is critically examined, in particular with regard to the reason given by Einstein for rejecting his theory. It is shown that the aspect Einstein found objectionable – the mutual dependence of the motions of particles when the (many-body) wavefunction factorises – is a generic attribute of his theory but that this feature may be removed by modifying Einstein's method in either of two ways: using a suggestion of Grommer or, in a physically important special case, using a simpler technique. It is emphasized though that the presence or absence of the interdependence property does not determine the acceptability of a trajectory theory. It is shown that there are other grounds for rejecting Einstein's theory (and the two modified theories), to do with its domain of applicability and compatibility with empirical predictions. That Einstein's reason for rejection is not *a priori* grounds for discarding a trajectory theory is demonstrated by reference to an alternative deterministic trajectory theory that displays similar particle interdependence yet is compatible with quantum predictions.

*Essay written in memory of J.T. Cushing.

## 1. INTRODUCTION

In 1927 Einstein presented for publication an article entitled "Does Schrödinger's Wave Mechanics Determine the Motion of a System Completely or Only in the Sense of Statistics?" in which he proposed a hidden-variable interpretation of quantum mechanics comprising a model of deterministic particle trajectories. At the last minute Einstein discovered what he believed was a fatal flaw in his theory and the piece was withdrawn. Although the contents of the manuscript have been described in the literature [1-5] (for many workers in the foundations of quantum mechanics their first acquaintance with it will have been the account in Jim Cushing's book *Quantum Mechanics: Historical Contingency and the Copenhagen Hegemony* [1]) there has been surprisingly little critical analysis of the technical details of Einstein's proposal. Commentary has tended to concentrate on its historical significance in Einstein's emerging analysis of wave-particle duality and quantum entanglement. In particular, it appears that Einstein's own assessment of his theory, and his reason for rejecting it, have been accepted at face value. Three key issues have been passed over: (a) whether the principle invoked by Einstein to exclude his theory is physically reasonable, (b) why Einstein's theory does not obey this principle, and whether it may be modified so that it does, and (c) whether there are other features of the theory that impinge on the case for it being regarded as a serious interpretation of quantum mechanics.

In investigating these questions, we shall attempt to elucidate aspects of Einstein's manuscript that have not been clearly represented in the literature. We find that the feature Einstein found objectionable – the mutual dependence of the motions of particles when the (many-body) wavefunction factorises – is indeed a generic feature of his theory but that, in a physically important special case, the "defect" may be removed by a simple modification of his method. Actually, a way to remove the "flaw" had already been proposed by Grommer, as mentioned by Einstein in his text, and we show that this variation also "works" in Einstein's sense. We write quotes here because, as we also emphasize, contrary to Einstein's opinion the presence or absence of the interdependence property does not determine the acceptability of a trajectory theory. Einstein's theory (and the two modified theories) *is* flawed, but for reasons not discussed by Einstein (or subsequent commentators); these have to do with the domain of applicability of the theory, and its relation with empirical predictions. That the reason given by Einstein for rejecting his approach is not *a priori* grounds for discarding a trajectory theory is illustrated by reference to an alternative deterministic trajectory theory that is compatible with the empirical content of quantum mechanics while displaying the same kind of particle coupling in product states abhorred by Einstein.

Note on notation: in Secs. 2-4 we stay close to Einstein's original notation, where particle coordinates are collectively denoted $q^\mu$, except that the summation convention for repeated indices is not assumed. In Sec. 5 we revert to more conventional notation where individual particle coordinates are denoted $\mathbf{x}_i$.

## 2. EINSTEIN'S TRAJECTORY THEORY AND ITS "FLAW"

The following summary of Einstein's paper draws upon Refs. [1-5], which are based on the original handwritten manuscript in the Einstein Archives (2-100).

Einstein considers a non-relativistic $N$ particle system whose configuration space (of dimension $n = 3N$) is labelled by coordinates $q^\mu$, $\mu = 1,...,n,$ and endowed with metric $g_{\mu\nu}$. On this space the stationary-state Schrödinger equation for a set of unit-mass particles is

$$E\psi = -\frac{\hbar^2}{2}\nabla^2\psi + V\psi, \qquad (1)$$

We assume that $\psi$ is the spatial factor in the full stationary-state wavefunction ($= \psi(q)e^{-iEt/\hbar}$). Einstein does not state that $\psi$ is real but it must be for various subsequent formulas, e.g., (2) below, to make sense. Einstein's idea is that the quantum expression for the quantity $E - V$ may be regarded as the kinetic energy associated with a set of $N$ point corpuscles with coordinates $q^\mu(t)$. He therefore identifies the two quantities:

$$\sum_{\mu,\nu} \tfrac{1}{2} g_{\mu\nu} \dot{q}^\mu \dot{q}^\nu = -\tfrac{1}{2}\hbar^2 \frac{\nabla^2\psi}{\psi}. \qquad (2)$$

Here $\nabla^2\psi = \sum_{\mu,\nu} g^{\mu\nu}\psi_{\mu\nu}$ where

$$\psi_{\mu\nu} = \frac{\partial^2 \psi}{\partial q^\mu \partial q^\nu} - \sum_\sigma \begin{Bmatrix} \sigma \\ \mu\nu \end{Bmatrix} \frac{\partial \psi}{\partial q^\sigma} \qquad (3)$$

and $\begin{Bmatrix} \sigma \\ \mu\nu \end{Bmatrix}$ are the Christoffel symbols built from $g_{\mu\nu}$.

The aim now is to infer from (2) expressions for the velocity components $\dot{q}^\mu$ in terms of the wavefunction. To achieve this, Einstein first passes to a new (in general, anholonmic) local frame of reference with respect to which each of the two expressions for the kinetic energy is diagonalized. He then assumes that the resulting summands may be equated term by term to obtain (up to a sign) expressions for the velocity components in the new frame. Finally, he passes back to the original frame to get the desired expressions for the $\dot{q}^\mu$s in terms of $\psi$. The construction of the frame, comprising a set of $n$ linearly independent orthonormal vector fields $A_{(a)}^\mu(q)$, $a = 1,...,n$, or $n$-tuple, follows the well-established procedure for finding the principal directions associated with a symmetric rank 2 tensor [6].

The required vectors are obtained (up to sign) by solving the equations

$$\sum_\nu \left( \psi_{\mu\nu} - \lambda_{(a)} g_{\mu\nu} \right) A_{(a)}^\nu = 0. \qquad (4)$$

These equations have non-trivial solutions if

$$\det\left( \psi_{\mu\nu} - \lambda_{(a)} g_{\mu\nu} \right) = 0 \qquad (5)$$

from which the eigenvalues $\lambda_{(a)}(q)$ may be found. The components of $g_{\mu\nu}$ and $\psi_{\mu\nu}$ with respect to the local frame are

$$g_{(ab)} = \sum_{\mu,\nu} g_{\mu\nu} A_{(a)}^\mu A_{(b)}^\nu = \delta_{ab}, \quad \psi_{(ab)} = \sum_{\mu,\nu} \psi_{\mu\nu} A_{(a)}^\mu A_{(b)}^\nu = \lambda_{(a)} \delta_{ab} \qquad (6)$$

so that, writing $\dot{q}^{(a)} = \sum_\mu \dot{q}^\mu A_\mu^{(a)}$, (2) becomes

$$\sum_a \tfrac{1}{2} \left( \dot{q}^{(a)} \right)^2 = -\sum_a \tfrac{1}{2} \hbar^2 \frac{\psi_{(aa)}}{\psi}. \qquad (7)$$

Both sides are now in the required diagonal form. Equating the summands and using (6) gives the $n$-tuple components of the velocity:

$$\dot{q}^{(a)} = \pm \hbar \sqrt{-\frac{\lambda_{(a)}}{\psi}}. \qquad (8)$$

Returning to the original system of coordinates, we have finally

$$\dot{q}^\mu = \sum_a \pm \hbar \sqrt{-\frac{\lambda_{(a)}}{\psi}} A_{(a)}^\mu \qquad (9)$$

where on the right-hand side we write $q^\mu = q^\mu(t)$. This is Einstein's proposed law of motion for the system point representing an individual *N*-body system associated with Schrödinger's equation (1). It is a "hidden-variable" theory in that the state of the system comprises quantities not contained in the wavefunction, namely, the particle positions ($\lambda_{(a)}$ and $A_{(a)}^\mu$ are functions of $\psi$ and are not hidden variables). The simultaneously well-defined particle velocities at each point are implied by (9) when the positions are specified. According to Einstein the sign ambiguity in (9) is to be expected in relation to quasiperiodic motions (it is suggested elsewhere in the manuscript that the motion may be determined completely by the requirement that the velocity changes only continuously). Einstein suggests that in this way one assigns uniquely to each Schrödinger solution motions of the particles, although it is not clear in what sense this association is unique.

In an added note Einstein refers to a calculation of Bothe that has led him to the realization that the scheme does not satisfy a general condition that he believes should be placed upon the law of motion of the system. He illustrates the objectionable feature by considering a two-body system whose Hamiltonian is an additive combination of Hamiltonians corresponding to the two systems. This requires that the configuration space metric and the Christoffel symbols each split up into two sets depending on one or the other particle coordinates. Thus

$$\nabla^2 \psi = \sum_{\mu,\nu=1,2,3} g^{\mu\nu} \psi_{\mu\nu} + \sum_{\mu,\nu=4,5,6} g^{\mu\nu} \psi_{\mu\nu}. \tag{10}$$

Then a possible solution of (1) is the factorised wavefunction

$$\psi(q) = \psi_1(q^1, q^2, q^3) \psi_2(q^4, q^5, q^6) \tag{11}$$

where $\psi_1$ and $\psi_2$ satisfy the respective one-body wave equations. Einstein demands that in this case the motions of the composite system are combinations of the individual particle motions. By this is presumably meant that the particle motions should be independent, each being calculable using solely the associated one-body wavefunction. But the law (9) does not have this property - it is easy to see that for the particle 1 component $\dot{q}^1$, say, $\lambda_{(a)}$ and $A_{(a)}^1$ generally depend on all the coordinates $q^1, ..., q^6$ for the wavefunction (11). (Moreover, the coupling between the particles does not generally fall off with interparticle distance.)

Einstein expresses this feature through the observation that for the state (11) $\psi_{\mu\nu} \neq 0$ when $\mu$ ($\nu$) is an index belonging to particle 1 (2) (for these indices $\psi_{\mu\nu} = \partial_\mu \psi_1 \partial_\nu \psi_2$ and so not all components can vanish for non-trivial solutions). Presumably the idea here is that the particle motions would be independent were all the components $\psi_{\mu\nu} = 0$ for this choice of indices. Although this last condition cannot be fulfilled, it is instructive to check this claim under the assumption that it can. Using this assumption, (5) implies that $\lambda_{(a)} = f_{(a)}(q^1, q^2, q^3)\psi$, $a = 1,2,3$, and $\lambda_{(a)} = f_{(a)}(q^4, q^5, q^6)\psi$, $a = 4,5,6$. Dividing by $\psi$, (4) then becomes

$$\sum_{\nu=1,2,3} \left( \frac{\psi_{1\mu\nu}}{\psi_1} - f_{(a)} g_{\mu\nu}(q^1, q^2, q^3) \right) A_{(a)}^\nu = 0, \quad \mu = 1,2,3 \tag{12a}$$

$$\sum_{\nu=4,5,6} \left( \frac{\psi_{2\mu\nu}}{\psi_2} - f_{(a)} g_{\mu\nu}(q^4, q^5, q^6) \right) A_{(a)}^\nu = 0, \quad \mu = 4,5,6. \tag{12b}$$

Consider the cases $a = 1,2,3$. The determinant of the matrix in (12a) is zero and the (therefore, non-trivial) solutions $A^\mu_{(a)}$, $\mu = 1,2,3$, are functions of just $q^1, q^2, q^3$. The determinant of the matrix in (12b) is nonzero and hence $A^\mu_{(a)} = 0$, $\mu = 4,5,6$. On the other hand, when $a = 4,5,6$, we find that $A^\mu_{(a)}$, $\mu = 4,5,6$, are functions of just $q^4, q^5, q^6$ and $A^\mu_{(a)} = 0$, $\mu = 1,2,3$. Inserting these results in (9) we find

$$\dot{q}^1 = \sum_{a=1,2,3} \pm \hbar \sqrt{-f_{(a)}} A^1_{(a)} \tag{13}$$

which is a function of only the particle 1 coordinates $q^1, q^2, q^3$. In the same way, $\dot{q}^4$, say, is a function only of $q^4, q^5, q^6$.

Einstein goes on to say that the preceding observation about $\psi_{\mu\nu}$ is connected with the fact that the $\lambda_{(a)}$s of the composite system do not coincide with the $\lambda_{(a)}$s associated with each of the individual systems.

It is worth emphasizing that Einstein is not querying the ability of many-body wave mechanics to express the notion of independence – it does that using the product state, the necessity of which Einstein had emphasized earlier [7]. Nor is he concerned with entangled quantum states. He focuses rather on what he regards as a failing of his own completion of wave mechanics, the mutual dependence of the hidden parameters in the case where quantum mechanics deems the systems to be independent. This matter is therefore to be distinguished from the later EPR analysis of nonproduct states, though of course the same kind of issue is at stake in both cases – interdependence in cases where Einstein requires there should be independence. To avoid confusion with the terminology used in connection with nonproduct quantum states (e.g., nonseparability, nonlocality, entanglement), in the sequel we shall speak of the mutual dependence, or coupling, of the hidden variables.

## 3. GROMMER'S SUGGESTION

At the end of the added note Einstein calls attention to a suggestion of Grommer that the perceived difficulty in the preceding scheme may be circumvented if $\psi$ is replaced by $\log\psi$ in the definition of the principal directions. He does not elaborate, remarking only that details would be given should the idea prove itself in examples. In fact, as we shall now see, Grommer's idea works in the sense Einstein would appreciate.

Making the substitution $\psi \to \log\psi$ in (4) we obtain a new set of eigenvalues and eigenvectors, $\kappa_{(a)}$ and $B^\mu_{(a)}$. In terms of the new function, the kinetic energy (2) becomes

$$\sum_{\mu,\nu} \tfrac{1}{2} g_{\mu\nu} \dot{q}^\mu \dot{q}^\nu = -\tfrac{1}{2} \hbar^2 \left( \nabla^2 \log\psi + (\nabla \log\psi)^2 \right). \tag{14}$$

With respect to the new local frame the two sides here are diagonalized:

$$\sum_a \tfrac{1}{2} \left( \dot{q}^{(a)} \right)^2 = -\sum_a \tfrac{1}{2} \hbar^2 \left( \kappa_{(a)} + (\log\psi)_{(a)}^2 \right) \tag{15}$$

where $(\log\psi)_{(a)} = \sum_{\mu} B^{\mu}_{(a)} \partial_{\mu} \log\psi$. Identifying summands and returning to the space coordinates, the revised law of motion is

$$\dot{q}^{\mu} = \sum_{a} \pm \hbar \sqrt{-\left(\kappa_{(a)} + (\log\psi)_{(a)}^{2}\right)} B^{\mu}_{(a)}. \tag{16}$$

Let us compute $\dot{q}^1$ for the product state (11). In this case the components $(\log\psi)_{\mu\nu} = 0$ when $\mu$ ($\nu$) is an index belonging to particle 1 (2) and the analogue of (5) ($\psi \to \log\psi$) implies that $\kappa_{(a)}$, $a = 1,2,3$, is a function of $q^1, q^2, q^3$ and $\kappa_{(a)}$, $a = 4,5,6$, is a function of $q^4, q^5, q^6$. The analogue of (4) thus becomes

$$\sum_{\nu=1,2,3} \left((\log\psi)_{\mu\nu}(q^1,q^2,q^3) - \kappa_{(a)} g_{\mu\nu}(q^1,q^2,q^3)\right) B^{\nu}_{(a)} = 0, \quad \mu = 1,2,3 \tag{17a}$$

$$\sum_{\nu=4,5,6} \left((\log\psi)_{\mu\nu}(q^4,q^5,q^6) - \kappa_{(a)} g_{\mu\nu}(q^4,q^5,q^6)\right) B^{\nu}_{(a)} = 0, \quad \mu = 4,5,6. \tag{17b}$$

Consider the cases $a = 1,2,3$. The determinant of the matrix in (17a) is zero and the (therefore, non-trivial) solutions $B^{\mu}_{(a)}$, $\mu = 1,2,3$, are functions of just $q^1, q^2, q^3$. The determinant of the matrix in (17b) is nonzero and hence $B^{\mu}_{(a)} = 0$, $\mu = 4,5,6$. On the other hand, when $a = 4,5,6$, we find that $B^{\mu}_{(a)}$, $\mu = 4,5,6$, are functions of just $q^4, q^5, q^6$ and $B^{\mu}_{(a)} = 0$, $\mu = 1,2,3$. We also have that $(\log\psi)_{(a)}$ is a function of $q^1,q^2,q^3$ ($q^4,q^5,q^6$) when $a = 1,2,3$ (4,5,6). Inserting all these results in (16) we find

$$\dot{q}^1 = \sum_{a=1,2,3} \pm \hbar \sqrt{-\left(\kappa_{(a)} + (\log\psi)_{(a)}^{2}\right)} B^{1}_{(a)} \tag{18}$$

which is a function of only the particle 1 coordinates $q^1, q^2, q^3$. In the same way, $\dot{q}^4$, say, is a function only of $q^4, q^5, q^6$.

Grommer's modification of Einstein's law thus has the feature Einstein desired, for arbitrary product states. From the available evidence it is not clear why Einstein did not pursue this variant of his theory.

## 4. CRITIQUE

Einstein seems to have regarded it as a principle of a trajectory theory that compatibility with the quantum notion of independence – expressed by a product wavefunction - requires that the corpuscles are physically independent. Before we examine whether particle coupling in such cases should be regarded as grounds for the automatic disqualification of a quantum trajectory theory, we enquire further into the reasons Einstein's theory has this property, and scrutinize other features of the theory bearing on its suitability to be afforded the status "interpretation of quantum mechanics".

One might suspect that the appearance of particle interdependence is an artefact of the use of a generalized configuration-space metric (subject to the decomposition (10)). However, the absence of this feature in Grommer's approach, developed within the same geometrical context, shows that this is not so. Indeed, in the case of the flat metric (19) discussed below, Einstein's law (9) still implies coupled particle motions. Given the generalized metric, the motivation for Einstein's introduction of the local frame eigenvectors – to diagonalize the kinetic energy - is clear. What is not clear is why Einstein opted for an arbitrary metric, and the consequent obligation to introduce the local frame, in the first place, given that this structure has no intrinsic relation with the quantum system of interest[1]. In fact, in an important special case the essential idea used by Einstein in deriving his law of motion can be expressed more simply without employing a local frame, as we now show. Moreover, this simpler theory enjoys the desirable independence property of Grommer's approach.

Suppose the configuration space is such that we can choose a global system of Cartesian coordinates:

$$g_{\mu\nu} = \delta_{\mu\nu} \tag{19}$$

(the usual case of many-body quantum mechanics). Then both sides of (2) are diagonalized,

$$\sum_\mu \tfrac{1}{2}(\dot{q}^\mu)^2 = -\sum_\mu \tfrac{1}{2}\hbar^2 \frac{\partial^2 \psi}{\psi \partial q^\mu \partial q^\mu}, \tag{20}$$

and we may seek, as Einstein did in the local frame, to equate the individual summands on each side. This gives for the velocity components

$$\dot{q}^\mu = \pm\hbar\sqrt{-\frac{\partial^2 \psi}{\psi \partial q^\mu \partial q^\mu}}, \quad \mu = 1,...,n. \tag{21}$$

This is now the law of motion of the individual system point. Consider a two-body system where $\psi$ factorises as in (11). Then it is clear that we get for the velocity of particle 1

$$\dot{q}^\mu = \pm\hbar\sqrt{-\frac{\partial^2 \psi_1}{\psi_1 \partial q^\mu \partial q^\mu}}, \quad \mu = 1,2,3, \tag{22}$$

which depends just on the particle 1 coordinates, and likewise for particle 2. Hence, using the law (21) the velocity of each particle is determined independent of the context, as Einstein requires. Einstein presumably missed this possibility because he developed his theory for metrics more general than (19). As noted above, for this metric his method still gives (9) for the law of motion and not (21).

The property of particle independence just demonstrated suggests that this model might be, on Einstein's reading, potentially acceptable (along with Grommer's version). Unfortunately, this approach, Grommer's proposal, and the original theory, are deficient in ways that render them unsuitable as serious candidate interpretations of quantum mechanics, at least in these undeveloped forms. The defects fall into two connected categories: the limited range of states and domains of configuration space to which the theories apply, and their failure to reproduce the empirical content of quantum mechanics.

---

[1] A clue may be that in this period Einstein was looking into unified field theories based on *n*-tuples [8].

Regarding the first category, the theories apply just to real stationary states for which $E \geq V$ (from (2)). Given the potential $V$, the latter restriction may be regarded as a condition on $E$ or on the configuration-space domain that the system may occupy. That is, we restrict attention either to states for which $E \geq V$ for all $q^\mu$ or, given the energy $E$, to the domain for which $q^\mu$ is restricted by $E \geq V$. These restrictions apply to all three theories. And that is not all: in the case of (9) and (18) the eigenvalues $\lambda_{(a)}$ and $\kappa_{(a)}$ must be real and simple (for the eigenvectors to form $n$-tuples), and the quantities $\lambda_{(a)}/\psi$ and $\left(\kappa_{(a)} + (\log\psi)_{(a)}^2\right)$ must be negative; in the case of (21) the term under the square root must be positive. These latter conditions imply further restrictions on the available portions of configuration space. The constraints on the energy or the domain are entirely artificial in that they do not correspond to any natural requirement of the quantum mechanical treatment.

To see how stringent these conditions are we examine the three contenders in the case of a space with metric (19) and a pair of one-dimensional harmonic oscillators of equal frequency $\omega$ ($V_1 = \tfrac{1}{2}\omega^2 q_1^2$, $V_2 = \tfrac{1}{2}\omega^2 q_4^2$) in the (product) ground state ($E_1 = E_2 = \tfrac{1}{2}\hbar\omega$):

$$\psi = A e^{-\omega q_1^2/2\hbar} e^{-\omega q_4^2/2\hbar}, \quad A = \text{constant.} \tag{23}$$

This state does not obey the condition $E \geq V$ for all $q_1, q_4$. Considering Einstein's proposal, we find for the eigenvalues, on solving (5),

$$\lambda_{(1)} = -\omega\psi/\hbar, \quad \lambda_{(4)} = \left(\psi/\hbar^2\right)\left(\omega^2 q_1^2 + \omega^2 q_4^2 - \hbar\omega\right). \tag{24}$$

Solving (4) the local orthonormal frame vectors are

$$A_{(1)} = \pm\left(q_1^2 + q_4^2\right)^{-1/2}(q_4, q_1), \quad A_{(4)} = \pm\left(q_1^2 + q_4^2\right)^{-1/2}(q_1, -q_4). \tag{25}$$

Substituting (24) and (25) into (9) we find, as expected, a dependence of each $\dot{q}^\mu$ on both coordinates. Eqs. (24) lead to the result that the velocities are real only in the elliptical domain $2V \leq E$, whereas of course the wavefunction extends over the whole $q_1 q_4$-plane.

In the case of Grommer's approach, we first note that for a general two-dimensional product state we get for the eigenvalues and eigenvectors (from (17a) and (17b))

$$\kappa_{(1)} = \partial^2 \log\psi_1/\partial q_1^2, \quad \kappa_{(4)} = \partial^2 \log\psi_2/\partial q_4^2 \tag{26}$$

$$B_{(1)} = \pm(1,0), \quad B_{(4)} = \pm(0,1). \tag{27}$$

Substituting these formulas in (18) we find that $\dot{q}^1$ coincides with the law (22), in the special case of two dimensions (and similarly for $\dot{q}^4$). Thus, for both theories we obtain for the oscillators

$$\dot{q}_1 = \pm\sqrt{\hbar\omega - \omega^2 q_1^2}, \quad \dot{q}_4 = \pm\sqrt{\hbar\omega - \omega^2 q_4^2}. \tag{28}$$

We thus obtain a pair of independent oscillators pursuing classical orbits. In this case the orbits are confined to the domain $V_1 \leq E_1$ and $V_2 \leq E_2$.

The origin of the problem of restricted applicability is Einstein's assumption, rather common in the history of quantum theory, that the quantity $E-V$ appearing in Schrödinger's equation may be regarded as a "kinetic energy". This assumption, which requires $E \geq V$, is known to lead to paradoxes [9]. It is, in fact, an arbitrary assumption, based on an unjustified allusion to classical mechanics. It is noteworthy that the development of a trajectory interpretation does not require this assumption. For example, in the de Broglie-Bohm approach the quantity that Einstein identifies as kinetic energy is regarded as quantum potential energy [1], and its sign imposes no restriction on the domain of applicability of the theory.

The other category of problems faced by the Einstein and related approaches is their compatibility with the empirical predictions of quantum mechanics. How, for example, are the trajectories connected with the outcomes of position measurements? In a 1927 letter Einstein maintains that his model "…can attribute quite definite movements to Schrödinger's wave mechanics, without any statistical interpretation…" [10] but surely *consistency* with the quantal probability formula has to be ensured. Indeed, at the Solvay conference later that year [11] Einstein was concerned precisely with the problem of interpreting $|\psi|^2$ and argued that a satisfactory interpretation requires that "…one does not only describe the process by the Schrödinger wave, but at the same time one localizes the particle during the propagation. I think that de Broglie is right in searching in this direction…". Regarding the capacity of each of the three theories to reproduce $|\psi|^2$ there are three options: (a) the flow of trajectories (obtained by varying the initial position in an ensemble of particles associated with the same wavefunction) generates $|\psi|^2$ directly (as in the de Broglie-Bohm theory [12]), or (b) the trajectories are used as building blocks to construct some other flow with this property (as in the model to be described in Sec. 5), or (c) the measurement process causes the system to jump into states so that the outcomes are distributed in accordance with $|\psi|^2$. We examine these possibilities in turn.

Einstein's trajectories (or those of the two modified theories) do not accord with option (a). This is obvious since, as we have seen above, in general the paths cannot navigate the entire region to which quantum mechanics attributes a finite probability of detection. A way of stating this is that the law (9) (or (16) or (21)) does not generate a conserved flow that maps the quantal distribution to itself:

$$\sum_\mu \nabla_\mu \left( |\psi|^2 \dot{q}^\mu \right) \neq 0. \tag{29}$$

This is so even in cases where $\dot{q}^\mu$ is defined for all relevant $q^\mu$. This significant defect may have become obvious if an extension to time-dependent, or at least complex, wavefunctions had been attempted (in the latter case one would take the real part of the right-hand side in (2)). Note that for real $\psi$ the conventional expression for the Schrödinger current vanishes. We can nevertheless entertain the possibility of a finite current $j^\mu$, as in Einstein's theory, but this must obey $\sum_\mu \nabla_\mu j^\mu = 0$ in order that the flow is compatible with $|\psi|^2$ [13].

The feature of restriction to subregions of configuration space also prevents the theories conforming to option (b): if the Einsteinian particles (or the other two sorts) are forbidden to pass into certain spatial regions to which quantum mechanics attributes a finite probability, an alternative flow cannot be constructed out of these particles that would reproduce $|\psi|^2$ in the entire configuration space.

Regarding option (c), it is unlikely the theories can be salvaged in this way. To be useful, the theories should apply to the measurement process. Strictly, this requires extending them to time-dependent wavefunctions but we need not attempt this for our purposes here. Suppose we consider a measurement process for which a stationary-state treatment is sufficient. Then, following our remarks above, we would find that the system trajectory (comprising object and measuring-device coordinates) is generally confined to subregions of the available configuration space where the total wavefunction is finite. Since the position of the system point will correspond to a particular measurement result, it follows that we cannot account for all possible results using a system point whose behaviour is determined by any of the stated laws.

Given that the flow is not one of the types (a)-(c) it is unclear what the achievement of Einstein's theory is, or how one should interpret the solutions of the equation of motion. Einstein regarded his construction as an analogue for Schrödinger's equation of the law of motion associated with the Hamilton-Jacobi equation in classical mechanics (which may give us a clue that he was interpreting the wavefunction as a formal computational device; recall that both the wavefunction and the classical Hamilton-Jacobi function are irreducibly defined on configuration space). However, Jacobi's particle equation follows uniquely from a specified dynamical transformation theory consistent with Newton's mechanics whereas its analogue (9) follows from the entirely *ad hoc* assumption (2).

To summarize, Einstein's hidden-variable theory applies just to a special case of quantum mechanics and the additional parameters do not play any useful role in accounting for measurement results. The physical meaning of the theory is therefore obscure. Einstein did not mention these potential defects of his theory. The same remarks apply to the two other versions of Einstein's idea considered above.

As regards the feature of his theory that Einstein did find flawed, it will be demonstrated next that his objection is unjustified by recalling an alternative deterministic trajectory theory that has the repellent feature of coupled particles in product states but is nevertheless compatible with quantum mechanics in that it is valid for arbitrary quantum states and the flow it defines generates (indirectly) the quantal probability [14]. We also bring out a further key property of Einstein's theory that it shares with the new theory.

## 5. A THEORY THAT DISPLAYS THE "EINSTEIN FLAW" YET IS COMPATIBLE WITH QUANTUM MECHANICS

### 5.1. Hidden-variable equations of motion

The model exploits the observation that Schrödinger's equation tacitly involves a (angular) degree of freedom that is not manifest in the conventional presentation of quantum mechanics (for a full account see [14]). The complex-valued *N*-body wavefunction $\psi(\mathbf{x}_1,...,\mathbf{x}_N)(=\psi_1+i\psi_2)$ may be represented by a 2-vector field $\psi_a \in \Re$, $a=1,2$, on the configuration space. The quantum formalism does not deal with the two vector components individually but rather defines physically relevant quantities in terms of "averages" over the internal index *a*. For instance, the usual expressions for the probability density and current for a set of particles of mass $m_i$, $i = 1,…,N$, may be written

$$\rho = |\psi|^2 = \sum_{a=1}^{2} \psi_a^2 \tag{30}$$

$$\mathbf{j}_i = \sum_{a,a'} (\hbar/m_i)\varepsilon_{aa'}\psi_a \nabla_i \psi_{a'}, \quad \varepsilon_{aa'} = -\varepsilon_{a'a}, \quad \varepsilon_{12} = 1. \tag{31}$$

Pursuing this observation, it can be shown that the spin 0 formalism may be treated formally as a special type of spin $\frac{1}{2}$ theory, with invariance group $SO(2)$ (corresponding to gauge transformations which maintain the realness of the 2-vectors). The aim is to make manifest the information encoded in the wavefunction that remains hidden if one deals only with the averages.

To proceed, it is easy to see that the spin 0 time-dependent Schrödinger equation,

$$i\hbar \frac{\partial \psi}{\partial t} = H\psi, \quad H = -\sum_{i=1}^{N} \frac{\hbar^2}{2m_i} \nabla_i^2 + V, \tag{32}$$

may be rewritten in 2x2 matrix form as follows:

$$i\hbar \frac{\partial \psi_a}{\partial t} = -\sum_{a'=1}^{2} H \sigma_{2aa'} \psi_{a'} \tag{33}$$

where $\sigma_2$ is a Pauli matrix. We next pass to a continuous representation of the wavefunction, replacing the discrete "spin" index $a$ by an (Euler) angle index $\alpha = (\alpha, \beta, \gamma)$. The wavefunction $\xi(\alpha)$ in this representation corresponding to $\psi_a$ in the discrete representation is given by

$$\xi(\mathbf{x}_1, ..., \mathbf{x}_N, \alpha) = \psi_1(\mathbf{x}_1, ..., \mathbf{x}_N) u_1(\alpha) + \psi_2(\mathbf{x}_1, ..., \mathbf{x}_N) u_2(\alpha) \tag{34}$$

where

$$u_1 = \left(2\sqrt{2}\pi\right)^{-1} \cos\frac{\alpha}{2} e^{-i(\gamma+\beta)/2}, \quad u_2 = -i\left(2\sqrt{2}\pi\right)^{-1} \sin\frac{\alpha}{2} e^{i(-\gamma+\beta)/2} \tag{35}$$

are basis functions corresponding to the basis column vectors into which $\psi_a$ can be expanded. Thus, while the coefficients $\psi_a$ are real, $\xi$ is in general complex. The Schrödinger equation (33) is now

$$i\hbar \frac{\partial \xi}{\partial t} = -(2/\hbar) H M_2 \xi \tag{36}$$

where $M_2(\alpha)$ is a component of the angular momentum differential operator $\mathbf{M}$ (transform of the Pauli matrices). The quantities (30) and (31) are calculated by averaging over the internal variable in its continuous rather than discrete guise:

$$|\psi|^2 = \int |\xi|^2 d\Omega \tag{37}$$

$$\mathbf{j}_i = \int \frac{i}{m_i} (\xi * M_2 \nabla_i \xi + \xi M_2 \nabla_i \xi*) d\Omega \tag{38}$$

where $d\Omega = \sin\alpha\, d\alpha\, d\beta\, d\gamma$.

The purpose of this exercise is that the continuous version of the Schrödinger equation (36) implies the following local conservation law for the quantity $|\xi|^2$:

$$\frac{\partial |\xi|^2}{\partial t} + \sum_{i=1}^{N} \nabla_i \cdot \left( |\xi|^2 \mathbf{v}_i \right) + (\mathbf{M}/-i\hbar) \cdot \left( |\xi|^2 \boldsymbol{\omega} \right) = 0 \tag{39}$$

where

$$\mathbf{v}_i = \left( i / m_i |\xi|^2 \right) \left( \xi^* M_2 \nabla_i \xi + \xi M_2 \nabla_i \xi^* \right) \tag{40}$$

and

$$\omega_1 = \omega_3 = 0, \quad (-\hbar/2)\omega_2 = \sum_{i=1}^{N} \frac{\hbar^2}{2 m_i} |\nabla_i \log \xi|^2 + V. \tag{41}$$

In itself the quantity $|\xi(\mathbf{x}_1,...,\mathbf{x}_N,\boldsymbol{\alpha})|^2$ has no significance in the quantum formalism (e.g., it is not gauge invariant) but here we shall interpret it as the probability density in the total external $(\mathbf{x}_1,...,\mathbf{x}_N)$ plus internal ($\boldsymbol{\alpha}$) configuration space. Integrating (39) over $\boldsymbol{\alpha}$ implies the usual probability conservation equation of quantum mechanics:

$$\frac{\partial |\psi|^2}{\partial t} + \sum_{i=1}^{N} \nabla_i \cdot \mathbf{j}_i = 0. \tag{42}$$

Hence the total probability is constant:

$$\frac{d}{dt} \int |\xi|^2 d\Omega \, d^3 x_1 ... d^3 x_N = 0. \tag{43}$$

A "hidden-variable" interpretation is obtained by introducing precise position variables associated with a set of particles and a further set of orientation variables associated with the whole system. Using the standard formulas connecting the angular velocity with the Euler angles, the correlated evolution of the variables is determined by the solution $\mathbf{x}_i = \mathbf{x}_i(\mathbf{x}_{i0},\boldsymbol{\alpha}_0,t), \boldsymbol{\alpha} = \boldsymbol{\alpha}(\mathbf{x}_{i0},\boldsymbol{\alpha}_0,t)$ to the coupled differential equations[2]

$$\frac{d\mathbf{x}_i}{dt} = \mathbf{v}_i(\mathbf{x}_1,...,\mathbf{x}_N,\boldsymbol{\alpha}) \tag{44}$$

$$\frac{d\boldsymbol{\alpha}}{dt} = (\sin\beta, \cos\beta \cot\alpha, -\cos\beta \csc\alpha) \omega_2(\mathbf{x}_1,...,\mathbf{x}_N,\boldsymbol{\alpha}). \tag{45}$$

Eq. (39) ensures that the deterministic flow obtained by varying continuously the initial conditions will map an initial density $|\xi_0|^2$ to $|\xi(t)|^2$ at any other time, for all quantum states. Regarding spatial positions alone, the model reproduces the quantal distribution via (37). We will not go into details but it is important to note that this theory is empirically sufficient, i.e., the stated properties of the model provide the basis for a theory of measurement that implies Born's rule for the relative frequency of measurement outcomes for all quantum observables.

---

[2] As with the de Broglie-Bohm theory [13] these identifications are unique only up to the possible addition of divergenceless fields to the currents in (39). There are obvious methodological similarities between this theory and de Broglie-Bohm (the latter corresponds to the mean flow). The detailed ontology is quite different, however.

**5.2. Particle interdependence in product states**

We come now to Einstein's abominable feature. For a two-body system, the quantum condition (11) for independence becomes in the present notation

$$\varphi(x_1, x_2) = \psi(x_1)\phi(x_2). \tag{46}$$

It is evident that factorisability into complex functions is equivalent to a kind of "entanglement" of the two sets of real fields corresponding to each one-body system that make up the two real fields associated with the total system:

$$(\psi_1 + i\psi_2)(\phi_1 + i\phi_2) = (\psi_1\phi_1 - \psi_2\phi_2) + i(\psi_1\phi_2 + \psi_2\phi_1). \tag{47}$$

The condition for factorisability in the 2-vector formalism therefore is:

$$\varphi_a = \sum_{a'a''} \phi_{a'}(\delta_{a1}\sigma_{3a'a''} + \delta_{a2}\sigma_{1a'a''})\psi_{a''}. \tag{48}$$

This product is readily translated into the continuous form. If $\xi$ and $\eta$ correspond to $\phi$ and $\psi$, respectively, the function corresponding to $\varphi$ is

$$\zeta(\mathbf{x}_1, \mathbf{x}_2, \boldsymbol{\alpha}) = \int \xi^*(\mathbf{x}_1, \boldsymbol{\alpha}') M_3(\boldsymbol{\alpha}') \eta(\mathbf{x}_2, \boldsymbol{\alpha}') d\Omega' \, u_1(\boldsymbol{\alpha}) + \int \xi^*(\mathbf{x}_1, \boldsymbol{\alpha}') M_1(\boldsymbol{\alpha}') \eta(\mathbf{x}_2, \boldsymbol{\alpha}') d\Omega' \, u_2(\boldsymbol{\alpha}) \tag{49}$$

where $M_1$ and $M_3$ are angular momentum differential operators.

It is evident that for the state (46) the velocity of, say, particle 1, given by (40) for $i = 1$, is generally an irreducible function of both sets of particle coordinates (and the angular parameters), regardless of how far apart the particles are. The coordinates of particle 2 do not drop out of the problem and there is, in general, no simple relation between the velocity computed from the total wavefunction $\varphi(\mathbf{x}_1, \mathbf{x}_2)$ and that from the one-body state $\psi(\mathbf{x}_1)$. In general, we cannot attribute to particle 1 its "own" wavefunction[3].

So, we can construct a theory with Einstein-like coupling of the particles whilst maintaining compatibility with the predictions of quantum mechanics. The coupling is hidden in that, averaging over the angle variables, we recover the usual factorised probabilities and velocities implied by (46):

$$|\psi(\mathbf{x}_1)|^2 |\phi(\mathbf{x}_2)|^2 = \int |\zeta(\mathbf{x}_1, \mathbf{x}_2, \boldsymbol{\alpha})|^2 d\Omega \tag{50}$$

---

[3] There are special cases of (46) where particle 1, say, is *relatively autonomous* in that $\mathbf{v}_1$ is independent of $\mathbf{x}_2$ whereas $\mathbf{v}_2$ depends on $\mathbf{x}_1$ and $\mathbf{x}_2$. In that case we may attribute to particle 1 its own wavefunction but not particle 2.

$$\frac{\nabla_i S_i}{m_i} = \frac{\int \mathbf{v}_i |\xi|^2 d\Omega}{\int |\xi|^2 d\Omega} \tag{51}$$

where $S_1$ ($S_2$) is the phase of $\psi$ ($\phi$).

A fundamental feature this theory shares with Einstein's (and the related theories of Secs. 3 and 4) is that the *context* of a particle is always relevant to its motion. Taken to its conclusion, this implies a primary role for the wavefunction of the Universe. In quantum mechanics one usually builds up the theory of complex systems starting from the simple one-body case, the tacit assumption having been made that there exist conditions under which a single body may be detached from its environment and studied in isolation. For this procedure to be valid the conditions have to be such that the wavefunction of the larger system, which is ultimately the wavefunction of the Universe, factorises into the relevant one-body function times a function depending on the variables of all the other systems. It is a feature of the present hidden-variable theory that factorisability expresses only the *statistical* independence of system 1 and its environment. Factorisability should not as a matter of course be afforded the stronger interpretation that the hidden-variable components of an individual ensemble element are independent. In this situation a complete physical account of particle 1 at the hidden-variable level generally involves the configuration of the hidden variables of the rest of the Universe. It follows that in (46) we must interpret the variables of "particle 2" as shorthand for "the rest of the Universe" (conceived here as a set of particles). The need to specify the configuration of all the particles in the Universe before we can compute the motion of a single particle is an issue of practicality rather than of consistency.

## 6. DISCUSSION

Einstein's trajectory theory and the two variations of it considered in Secs. 3 and 4 apply only to a limited set of wavefunctions, and for each of those generally a limited region of configuration space. The theories are unable to reproduce the quantum statistical results of position measurements. Einstein rejected his theory not for these reasons but because factorisability of the wavefunction is not translated into decomposition of the force acting on the particles. We have seen by the explicit construction of a viable alternative that in itself this is not a good reason to reject a trajectory theory.

In connection with Einstein's independence requirement, it may be that he was translating into the quantum realm a property of the corresponding classical system when the Hamiltonian decomposes. We note though that there is a fundamental difference between the way forces act in these sorts of quantum trajectory theories and in classical mechanics. Failure to appreciate this in, e.g., the de Broglie-Bohm theory has led to some confusion. The key point is that the trajectories of a many-body quantum system are correlated not because the particles exert a direct force on one another (*à la* Coulomb) but because all are acted upon by an entity – mathematically described by the wavefunction or functions of it – that lies beyond them. Although the wave is a function of all the particle coordinates, it is not *produced* by them[4]. This holistic notion of action is not present in or anticipated by the classical paradigm. Hence, in the special case of a decomposable Hamiltonian and factorised wavefunction we cannot invoke classical intuition and infer that the particles will behave independently in all such theories (the de Broglie-Bohm approach is an example of a theory that does have this property). In this connection, theories of the type pioneered by Einstein pose further as yet uncharted questions (complementary to those associated with Bell's theorem) to do with the degree of interconnectedness of the hidden variables that would be compatible with quantum independence.

---

[4] A way to treat particle back-reaction in de Broglie-Bohm theory is described in [15].

As became clear in the subsequent EPR story, Einstein [16] felt that physics would become impossible if the mutual dependence of separated objects occupying distinct regions of space was admitted as a general property of matter, for it would deny the possibility of studying segments of matter in isolation and the subject would lose its empirical basis. The idea needs to be examined in greater detail than is possible here but, reflecting on this question in the context of the model presented in Sec. 5, it seems that there is nothing logically objectionable in the notion that all objects in the Universe are in fact mutually dependent, all being involved in the behaviour of a single particle even when this is separated to a large distance, and for factorisable or nonfactorisable wavefunctions. As we have seen, such linkages do not prevent us from defining what we mean by an individual system, and we can ask and answer meaningful empirical questions that test the theory. Indeed, such a theory need not contradict any prediction of quantum mechanics since the testable aspects may be independent of the details of the complex substructure. Rather than spelling the end of science, specification of the conditions under which systems may be regarded as independent, or will be revealed as such in experiment, simply becomes a scientific problem in its own right. In his 1927 manuscript and later Einstein wanted rather to impose the conditions for independence on theory but, to adapt an argument used by him in a different context, it is more appropriate to say that it is *theory* that tells us when it is legitimate to regard objects as independent, and which aspects of that independence we may observe.